\documentclass[aps,prl,twocolumn,showpacs,superscriptaddress,groupedaddress]{revtex4}  
\usepackage{graphicx}  
\usepackage{dcolumn}   
\usepackage{bm}        
\usepackage{amssymb}   

\begin{document}

\widetext
%
%
\title{Transconductance fluctuations as a probe for interaction induced quantum Hall states in graphene}

\author{Dong Su Lee}
\affiliation{Max-Planck-Institut f\"{u}r Festk\"{o}perforschung,
Heisenbergstr. 1, D-70569 Stuttgart, Germany}
\author{Viera Sk\'{a}kalov\'{a}}
\affiliation{Max-Planck-Institut f\"{u}r Festk\"{o}perforschung,
Heisenbergstr. 1, D-70569 Stuttgart, Germany}
\author{R. Thomas Weitz}
\affiliation{Max-Planck-Institut f\"{u}r Festk\"{o}perforschung,
Heisenbergstr. 1, D-70569 Stuttgart, Germany}
\author{Klaus von Klitzing}
\affiliation{Max-Planck-Institut f\"{u}r Festk\"{o}perforschung,
Heisenbergstr. 1, D-70569 Stuttgart, Germany}
\author{Jurgen H. Smet}
\affiliation{Max-Planck-Institut f\"{u}r Festk\"{o}perforschung,
Heisenbergstr. 1, D-70569 Stuttgart, Germany}
\date{\today}

\begin{abstract}
Transport measurements normally provide a macroscopic, averaged view of the sample, so that disorder prevents the observation of fragile interaction induced states. Here, we demonstrate that transconductance fluctuations in a graphene field effect transistor reflect charge localization phenomena on the nanometer scale due to the formation of a dot network which forms near incompressible quantum states. These fluctuations give access to fragile broken-symmetry and fractional quantum Hall states even though these states remain hidden in conventional magnetotransport quantities.
\end{abstract}

\maketitle

%
Progress in graphene sample quality starts to disclose rich physics related to Coulomb interactions as well as interaction induced lifting of symmetries associated with the spin and pseudospin degrees of freedom. This is visible in macroscopic magnetotransport studies as additional incompressible quantum Hall condensates. If interactions are ignored, quantum Hall plateaus are expected
and seen in experiment for the filling factor sequence ${\nu} = \pm2, \pm6, \pm10, \cdot\cdot\cdot$
due to the Dirac fermion nature of the particles and the initial SU(4) symmetry~\cite{Novoselov2005, Zhang2005}. When disorder becomes weaker other more fragile QH states induced by interactions emerge.
For instance spontaneous symmetry breaking gives rise to the states with SU(2)
symmetry at ${\nu} = 0, \pm4, \pm8, \cdot\cdot\cdot$ as well as the fully broken symmetry states at ${\nu} = \pm1, \pm3, \cdot\cdot\cdot$~\cite{Zhang2006, Goerbig2006, Yang2007, Gusynin2008, Checkelsky2008, Jung2009}.
Also fractional quantum Hall states appear~\cite{Toke2007, Papic2009, Bolotin2009, Du2009, Dean2011}.

A key requirement to observe these fragile states in transport has so far been the fabrication of better quality samples. This has been accomplished by placing graphene on BN~\cite{Dean2010, Dean2011}, or by suspending graphene~\cite{Bolotin2009, Du2009}. In the quest for observing fragile or novel incompressible states, probing a smaller area may be an alternative to circumvent the challenges of producing even higher mobility samples, since the sample may be much cleaner on the nanometer scale~\cite{Li2007, Miller2009, Song2010, Jung2011}. However, to obtain such microscopic information we are usually forced to resort to sophisticated local probe methods which prevent the blurring by disorder induced averaging. Indeed, transport studies normally offer an averaged view of the sample only and this averaging masks the interaction induced physics.

That said, we note that graphene magnetotransport traces are frequently cluttered with fluctuations. Here we demonstrate that these fluctuations reflect processes that occur very locally. They are a manifestation of charge localization and are more pronounced in the transconductance. A systematic study allows observing higher order fractional quantum Hall and broken symmetry states even if these do not show up as quantized states in the Hall or longitudinal resistance traces. They make phenomena on the nanometer scale visible in a macrocopic transport experiment despite significant disorder.

Our studies were carried out on both supported and suspended exfoliated graphene in either a two-terminal or four-terminal configuration. The representative data discussed here stem from a suspended two-terminal device shown in Fig.~1. Fabrication procedures followed closely those previously reported~\cite{Bolotin2009, Du2009}. The device had a width of 3~$\mu$m and was about 0.8~$\mu$m long. A dc source-drain voltage was applied. The density was tuned with the doped Si substrate separated from the graphene device by a 150 nm thick SiO$_2$ layer and an equally thick air gap. Charge neutrality was reached at a gate voltage $V_{\mathrm{bg}}$ of about 1.2~V. The density inhomogeneity, $\Delta n_{\mathrm d}$ $\sim$ 6 $\times$ 10$^{10}$ cm$^{-2}$, was determined from the full-width at half-maximum of the field effect resistance (not shown).

Figure~1(a) plots the two-terminal conductance in units of $e^2/h$ as a function of the density for perpendicular fields from 1~T up to 15~T in 1~T steps. There are no surprises here. Plateaus or precursors of a plateau are observed for the integer states at filling $\nu = \pm 2, \pm 6, \cdot\cdot\cdot$ as expected for monolayer graphene. Also broken symmetry states at $\nu = 0$ and $\nu = \pm 1$ can be discerned. They have been attributed to the lifting of the spin and/or valley degeneracy of the Landau levels. Other broken symmetry states are missing and also fractional states remain hidden. Figure~1(b) displays the single trace recorded at 15~T for closer inspection. It is not your textbook quantum Hall curve and looks noisy. We are not dealing with noise though. These fluctuations are amplified in the transconductance $g_{\mathrm m}$. When applying an additional ac modulation ${\delta}V_{\mathrm{bg}}$ to the backgate, an oscillating source-drain current ${\delta}I_{\mathrm{ds}}$ is induced and $g_{\mathrm m} = {\delta}I_{\mathrm{ds}}/{\delta}V_{\mathrm{bg}}$. The frequency of ${\delta}V_{\mathrm{bg}}$ was 433 Hz and its root-mean-square amplitude was 10~mV. This corresponds to a density change of $5.8 \times 10^{8} {\rm{cm}}^{-2}$ or the average addition of a single electron to an area with 0.5~${\mu}$m diameter. In the transconductance, the background has vanished and only the fluctuations are left over in Fig.~1(c). This greatly simplifies a systematic study of these fluctuations.

Repeating this experiment for different $B$-values yields the color rendition of Fig.~2(a). The rich set of features can be classified into several groups of parallel lines. For instance at the center, we find vertical lines parallel to the $\nu = 0$-line. A bundle of lines parallel to $\nu = -2$ and $\nu = 2$ appears left and right. At the bottom, lines with a slope identical to the $\nu = \pm 6$-line can be discovered. These lines parallel to filling factor slopes at which the system is expected to condense in an incompressible state are reminiscent of the spikes that have been observed in local compressibility measurements using a single electron transistor on graphene ~\cite{Martin2009} as well as of earlier local compressibility data on GaAs two-dimensional electron systems~\cite{Ilani2004}. They were attributed to non-linear screening and charge localization in a landscape of compressible dots separated by an incompressible sea, which forms as we approach complete filling of a Landau level. Since this localization picture is crucial for understanding Fig.~2(a), it is described here.

A widespread view of localization in the quantum Hall regime is based on the association of localized states with equipotential lines which either enclose potential hills or are trapped inside a valley of the disorder potential. Plotting the position in the density versus $B$-field plane at which such localized states get filled, would generally result in curved, non-monotonic lines that may cross as they change their shape and move up or down in the disorder landscape when changing $B$ to maintain a fixed number of enclosed flux quanta. None of these properties are compatible with the observed parallel lines. This picture only holds in the strong disorder limit when single particle physics prevails~\cite{Ilani2004}.

Localization proceeds in a different manner as illustrated in Fig.~3(a). Screening and Coulomb blockade physics are key ingredients. Charge carriers will redistribute and generate a spatially dependent density profile in an attempt to flatten the bare disorder potential. As long as the local density of states is not exhausted, they are able to accomplish this task. As a Landau level approaches complete filling in some regions the required density to flatten the bare disorder potential locally exceeds the level degeneracy. Hence, the bare disorder potential can not be screened away. Here a potential barrier emerges. Eventually, compressible regions or dots are enclosed by an incompressible lake, Coulomb blockade physics sets in. The discrete nature of charge becomes relevant. Elementary charges can be added to the remaining compressible dots only one at a time when the overall density is raised sufficiently. Each charging event corresponds to filling a localized state. As the average density is increased further, more and more charges are added to the dots. The same landscape of compressible areas appears and the same charging physics recurs at a higher $B$, but at the same density deviation from complete filling. The filling of specific localized states therefore evolves along lines parallel to each other and to the corresponding integer filling factor in the $(n,B)$-plane as schematically drawn in Fig.~3(a). The same holds for fractional states but with fractional charges~\cite{Martin2004}.

It is natural to attribute the parallel lines in the transconductance to charge localization physics in this network of dots that forms when charge carriers are locally unable to screen the bare disorder. Current may flow from the source to drain through this network as well as extended edge states (inset Fig.~3(c)). During a backgate modulation cycle, the Coulomb blockade may be lifted for some dots or activated for others and transport channels are turned on, off or altered. This results in an oscillating source-drain current. At a different $B$, the same transport channels would appear and disappear at the same deviation from complete filling. Hence, a plot of the transconductance in the $(n,B)$-plane should be crowded by sets of parallel lines running along filling factors at which incompressible ground states form much the same way as in local compressibility measurements~\cite{Martin2009}. This is indeed confirmed in the experiment of Fig.~2(a). Weak line features were previously also observed in GaAs 2D electron systems~\cite{Cobden1999} and graphene~\cite{Branchaud2010, Velasco2010} when just measuring the resistance. They were connected to single particle integer quantum Hall states only. The transconductance here reveals the much more fragile broken symmetry states and fractional quantum Hall states in graphene. For instance, lines parallel to $\nu = 0$, $\nu = 1$ and $\nu = -1/3$ are easily discovered without further data processing in Fig.~2(a). Additional incompressible states can be identified by using a numerical analysis. It is based on the calculation of the correlation function $C(\nu)$ $=$ ${\sum}(D_{kl} D_{pq} {\delta}_{\nu})$ for the discrete data set. The calculation is illustrated in Fig.~3(b). The data form a matrix and $D_{ij}$ corresponds to the data point at density $n_{i}$ and field $B_{j}$. The product of any two data points $D_{kl}$ and $D_{pq}$ contributes to the sum in $C(\nu)$ provided they fall within a band in the $(n,B)$-plane bordered by two lines with a slope corresponding to filling $\nu$ and with a width of two pixels. In this case $\delta_{\nu}$ equals one. Otherwise, $\delta_{\nu}$ equals zero and this pair of points does not add to the sum. $C(\nu)$ is evaluated as a function of $\nu$ and the result is plotted in a polar diagram as shown in Fig.~2(b)$-$(c). The angle corresponds to the filling. To reduce the computational effort the calculation of $C(\nu)$ is restricted to a limited rectangular window in the $(n,B)$-plane. The center of the analysis window is placed in areas where interesting features are expected. Key locations have been marked in Fig.~2(a).
$C(\nu)$ for different analysis windows are overlaid, but colored differently. In Fig.~2(b) maxima of $C(\nu)$ appear at $\nu = -2, -4, -6$ and $-10$. Note that the $\nu = -4$ quantum Hall state, attributed to a symmetry reduction from SU(4) to SU(2), did not show up in the conductance up to 15~T. In the color rendition lines for other broken-symmetry states at ${\nu}$ $=$ 0 and ${\nu}$ $=$ $\pm$1 are seen down to $\sim$0.5~T and $\sim$1~T, respectively while in the two-terminal conductance these broken-symmetry states give rise to plateaus or minima only above 4~T. Fractional quantum Hall states at filling factors $\nu$ $=$ $-$1/3, $-$2/3 are clearly seen in the $C(\nu)$ spectra of Fig.~2(c). They are the prominent fractional states observed in the cleanest of all graphene flakes~\cite{Bolotin2009, Du2009}. But also the higher order ${\nu}= -2/5$ fractional state is present in Fig.~2(b) as predicted in Ref.~\cite{Toke2007, Papic2009} and as recently reported in local compressibility studies~\cite{Feldman2012}. A small feature is seen near $\nu = 1/2$, but more work is needed to firmly establish its significance. None of the observed fractional quantum Hall states appeared in the conductance.
The appearance of parallel lines in the transconductance serves as indirect evidence for screening governed localization and the emergence of a dot network. In graphene more direct evidence can be obtained. Under normal quantum Hall conditions the transport through this network is short-circuited by the extended edge channels. They prevent a measurement on only this network. However, graphene exhibits a true insulating quantum Hall state near $\nu = 0$ due to valley symmetry breaking of the zero energy mode~\cite{Yang2007, Gusynin2008, Checkelsky2008, Jung2009}. In this regime, the chemical potential does not cross any of the valley split Landau levels even at the edge where they bend up or down. An inset of Fig.~3(c) displays the level diagram. Source-to-drain current flow then can only proceed by charge tunneling through the dot network. The differential conductance, $dI_{\mathrm{ds}}/dV_{\mathrm{ds}}$, as a function of $V_{\mathrm{bg}}$ is plotted at the top of Fig.~3(c). At the border of the insulating regime, transport features are present. Deep into the $\nu = 0$ state, the system is too incompressible and no current can flow. The bottom panel of Fig.~3(c) displays the differential conductance as a function of $V_{\mathrm{bg}}$ and $V_{\mathrm{ds}}$. It resembles a stability diagram and exhibits a series of Coulomb diamonds. This confirms the existence of a network of dots. From the diamonds it is possible to extract typical dot sizes. They range from 160 to 400 nm and reflect the disorder length scale. The conductivity near the charge neutral point at zero field is approximately $3.5 e^2/h$. From this value, we can deduce a mean free path $l_{\mathrm {mfp}}$ $\sim$ 350 nm~\cite{Mucciolo2010}, which is consistent with the length scale estimated from the Coulomb diamonds. The irregularity of the Coulomb blockade oscillations in $dI_{\mathrm{ds}}/dV_{\mathrm{ds}}$ is attributed to the network character or to the irregular shape of the QDs for which chaotic resonances have been predicted~\cite{Bardarson2009}. We note that signatures of such a complex network of QDs in graphene have recently also been reported in scanning tunneling microscopy~\cite{Jung2011}.

The transconductance fluctuations in Fig.~2(a) allow the extraction of a lower bound for the density variation across the sample. When filling localized states produces features parallel to a nearby quantum Hall filling factor $\nu_{\rm i}$, it confirms a posteriori that the landscape of dots is independent of the $B$-field and the assumption of a fixed bare disorder potential is correct. It also implies that the localized states associated with quantum Hall state $\nu_{i}$ all get filled within a fixed density band whose width ${\Delta}n_{\rm d}$ is determined by the difference between the density extrema $n_{\rm max}$ and $n_{\rm min}$ of the spatial density variation that emerges to screen the bare disorder potential. In the large $B$-field limit, these bands of localized states are well separated and their width immediately yields the maximum density variation across the sample. Here, charging lines running parallel to adjacent quantum Hall fillings $\nu_{\rm H}$ and $\nu_{\rm L}$ still overlap. For instance, features parallel to $\nu_{\rm H} = -2$ are seen beyond the $\nu_{\rm L}=-1$-slope. Hence, the density variation exceeds $B/\phi_{0}\cdot (\nu_{\rm H}- \nu_{\rm L})$ even up to 15~T. This case is illustrated in Fig.~3(a). This sets the lower bound for ${\Delta}n_{\rm d}$ at $3.6 \times 10^{11} {\mathrm{cm}}^{-2}$. This lower limit is much larger compared with the disorder estimate from the field effect curve $(6 \times 10^{10} \mathrm{cm}^{-2})$. The latter presumably reflects the density variance rather than the difference between the extrema.

We conclude that transconductance fluctuations offer a window to fragile incompressible quantum Hall states, because they reflect
charging effects on the nanometer scale due to the appearance of the gap in the energy spectrum even if the disorder is
so large that the quantization signatures of these Hall states are entirely missing in standard magnetotransport. The method described here is an attractive alternative in the quest for observing unconventional quantum Hall
states.

We acknowledge financial from the German Science Foundation (SPP1459).
We thank G.E. Kim for fruitful discussions and A. G\"{u}th, T. Reindl and M. Hagel for sample preparation.

\clearpage

\begin{figure*}
\begin{center}
\includegraphics[width=176mm]{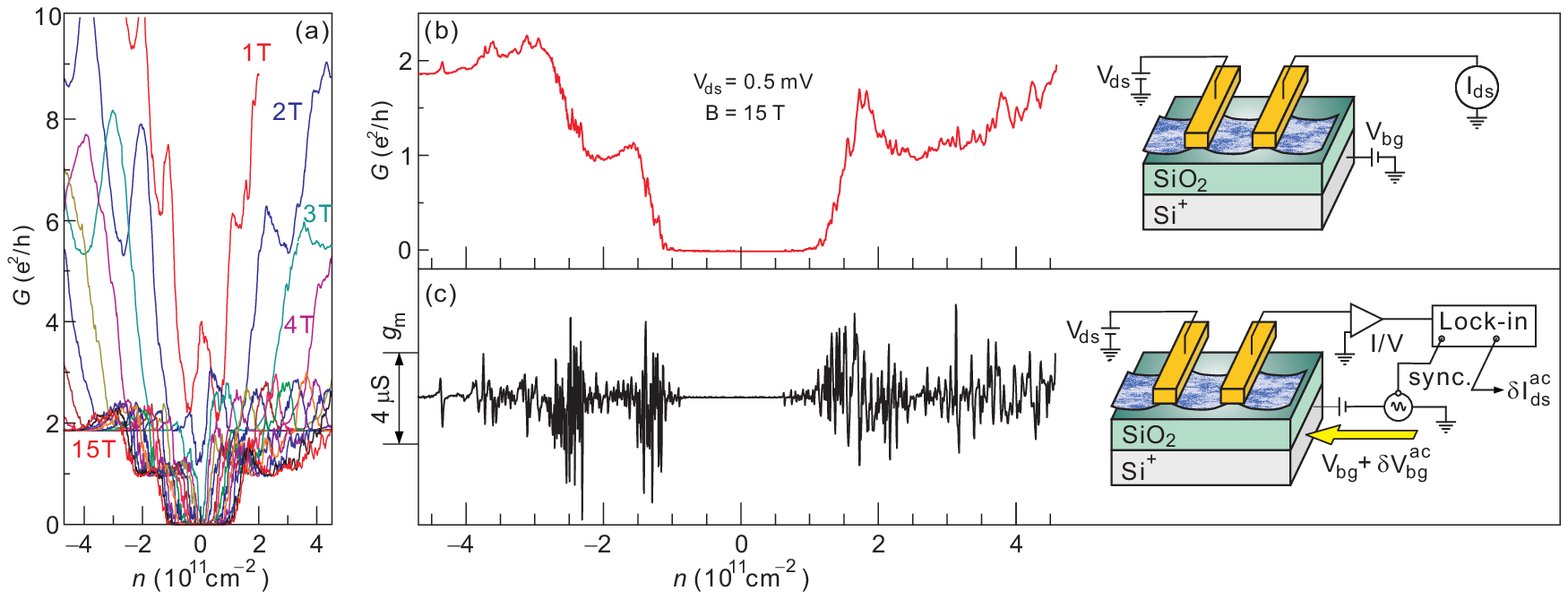}
\end{center}
\caption{(a) Two-terminal conductance as a function of density at fields
from 1~T to 15~T in 1~T steps. (b)$-$(c) Comparison of conductance (b) and transconductance (c)
curves measured at $B =$ 15~T. Both measurements were done with a dc-voltage bias of $V_{\mathrm {ds}} =$ 500 ${\mu}$V.
Insets show the schematics of the measurements.
} \label{figure1}
\end{figure*}

\begin{figure*}
\begin{center}
\includegraphics[width=171.453mm]{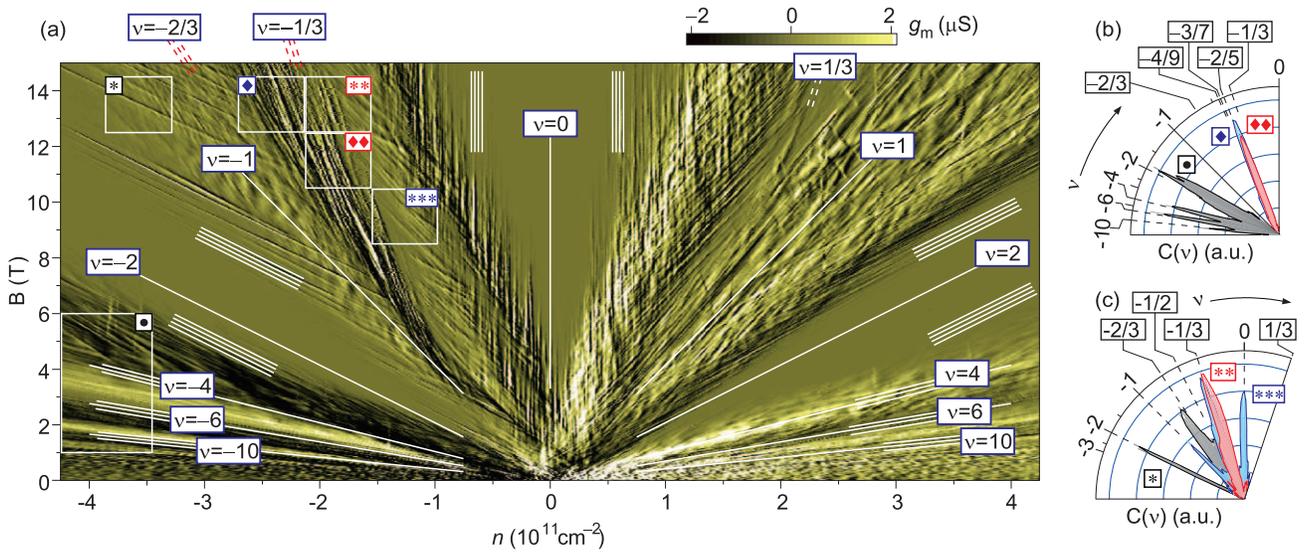}
\end{center}
\caption{(a) Color rendition of the transconductance in the $(n,B)-$plane. (b)$-$(c) Correlation function C obtained by analyzing the data in the windows shown in (a). The details of this analysis method are shown in Fig.~3.
} \label{figure2}
\end{figure*}

\begin{figure}
\begin{center}
\includegraphics[width=73.588mm]{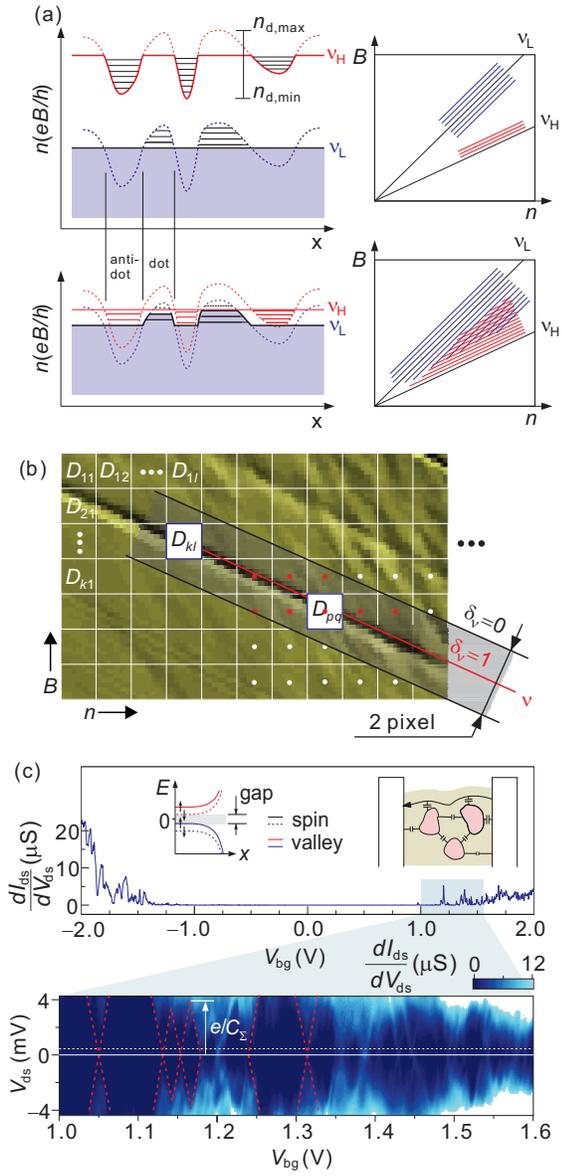}
\end{center}
\caption{(a) Schematic illustration of the formation of QDs and the development of the compressible spikes in the $(n-B)$-plane.
(b) Schematic illustration of the numerical analysis to obtain the spectra shown in Fig.~2. (c) Differential conductance and stability diagram measured at 1.4~K and 14~T. Inset shows an energy diagram along the sample on the left and a schematic of the network of of compressible QDs with extended edge states on the right.
} \label{figure3}
\end{figure}

\end{document}